\documentclass[twoside]{article}

\pdfoutput=1

\usepackage{graphicx}

\usepackage{cmpj2e}

\title[Computer simulations of ultrathin water film]%
{Molecular dynamics simulations of ultrathin water film confined between flat diamond plates}

\author[A.V. Khomenko, N.V. Prodanov]{A.V. Khomenko, N.V. Prodanov}

\address{Sumy State University, 2 Rimskiy-Korsakov Str., 40007 Sumy, Ukraine}

\begin{document}

\maketitle

\begin{abstract}
Molecular dynamics simulations of ultrathin water film confined between atomically flat rigid diamond plates are described. Films with thickness of one and two molecular diameters are concerned and TIP4P model is used for water molecules. Dynamical and equilibrium characteristics of the system for different values of the external load and shear force are investigated. An increase of the external load causes the transition of the film to a solidlike state. This is manifested in the decreasing of the diffusion constant and in the ordering of the liquid molecules into quasidiscrete layers. For two-layer film under high loads molecules also become ordered parallel to the surfaces. Time dependencies of the friction force and the changes of its average value with the load are obtained. In general, the behaviour of the studied model is consistent with the experimental results obtained for simple liquids with spherical molecules.
\pacs 61.20.Ja, 61.20.Ne, 61.30.Hn, 68.35.Af
\end{abstract}

\section{Introduction}

Studying of friction on the atomic level is important from fundamental and practical sides. The development of nanotechnology and the creation of new devices on its base require a deep understanding of the nanoscale processes leading to wear and friction \cite{Braun2006,Bhush2005,Dedko2000,Yoshi1993}. In the nanotribological studies there is a phenomenon which has a special place and is called boundary lubrication. It corresponds to the situation when rubbing surfaces are separated by an ultrathin (a few molecular diameters) lubricant film. Boundary lubrication is important not only for the nanotechnology. It is observed under the ordinary conditions, because almost always there is a lubricant between the solids (called ``the third bodies'' by tribologists) --
either a specially chosen lubricant film, or wear debris produced by sliding, or water adsorbed from air, etc. Boundary lubrication is also important in the macromachines at stop/start moments, when the lubricant is squeezed out from the contact area and the surfaces come into direct contact \cite{Braun2006,Bhush2005,Dedko2000,Yoshi1993}.

As show experiments with surface-force apparatus (SFA) and the computer simulations the behaviour of ultrathin liquid film confined between two solid substrates entirely differs from the bulk one. The main features of a boundary lubricant are \cite{Braun2006,Yoshi1993,GeeML1990,GaoJ1997}:
\begin{enumerate}
\item Liquid molecules confined between two atomically flat surfaces become more ordered and tend to organize quasidiscrete layers, where the average local density of the liquid oscillates with distance normal to the boundaries.
\item The mobility of molecules in confined films considerably decreases in comparison with bulk liquids. This is manifested in the decrease of the diffusion constant and in the increase of the viscosity and molecular relaxation times. The ``effective'' viscosity can be $10^{5}$ times larger and relaxation times can be $10^{10}$ times slower than  the bulk value.
\item Boundary lubricant may exhibit two different responses in shear: a liquidlike, in which the liquid responds to the deformation by flow, and a solidlike characterized by observation of the development of yield stress in the confined fluid. In computer simulations with simple Lennard-Jones (LJ) liquids between atomically structured surfaces the existence of abrupt liquid to solid transitions in films thinner than 6 molecular diameters was predicted, with the molecules becoming ordered both perpendicular and parallel to the surfaces. However, the horizontal order disappears for unstructured surfaces. In confined fluids both positional and orientational orderings are caused not only by mutual interactions of liquid molecules but also by the presence of two solid surfaces close together.
\item For velocities of moving substrate less than some critical value the stick-slip motion may be observed. The stick-slip regime is characterized by intermittent stops (stick) and slips.
\end{enumerate}

Analytical models are widely used for explaining the experimental results. But these models are applicable only in special cases and, as a rule, they are only qualitative \cite{Carls1996,Popov2001,Arans2002,Khome2003}. Computer simulations offer an additional tool for studying of the atomic-scale friction and wear. In particular, molecular dynamics (MD) simulations greatly help in the understanding of the nanotribological processes \cite{GaoJ1997,HeoSJ2005,Thomp1990,Thomp1992,Braun2001}.

In this article the MD simulations of ultrathin water film confined between atomically flat diamond plates are described. The choice of water as a liquid for modelling is motivated by its importance, ubiquity and unique, in particular tribological properties \cite{GeeML1990,Israe1998}. An introducing of a water monolayer (only 0.25 nm thick) between atomically flat mica surfaces of the SFA causes the decrease of friction more than by the order of magnitude \cite{Ruths2005}. The effectiveness of a water monolayer to lower the friction force is attributed to the ``hydrophilicity'' of the mica surface (mica is ``wetted'' by water) and to the existence of a strongly repulsive short-range hydration force between such surfaces in aqueous solutions, which effectively removes the adhesion-controlled contribution to the friction force. Under the described conditions the friction force obeys the first Amontons' law which states that frictional force is proportional to the external load.

The objective of the present work is to investigate the behaviour of the water film consisting of TIP4P molecules confined between totally rigid flat diamond plates and to check the model for consistency with the experiments. This in turn may give us estimates of the reliability of using the TIP4P model for water molecules, and the applicability of the approximation of absolutely rigid plates in the modelling of the tribological systems.

\section{Model}

The simulations are performed in a planar, Couette geometry that closely resembles the experimental systems \cite{Yoshi1993,GeeML1990}. A thin film of water molecules is confined between two solid walls with periodic boundary conditions in the plane of the walls. Each wall consists of 1152 atoms forming two (001) crystallographic planes of diamond lattice. Although the experimental study of friction of the diamond plates separated by the ultrathin water film has not been carried out yet but the diamond is transparent for the light and therefore can be used in the SFA. The walls are considered as absolutely rigid and the elasticity of plates is not included into the model. But taking into account that diamond is one of the hardest materials we decided to verify such approximation. The water films of one and two molecular diameters thickness have been investigated. At the beginning of simulations one water layer consists of 196 molecules. The maximum value of particles involved in the simulations is 2696.

\begin{figure}[htb]
\centerline{\includegraphics[width=0.49\textwidth]{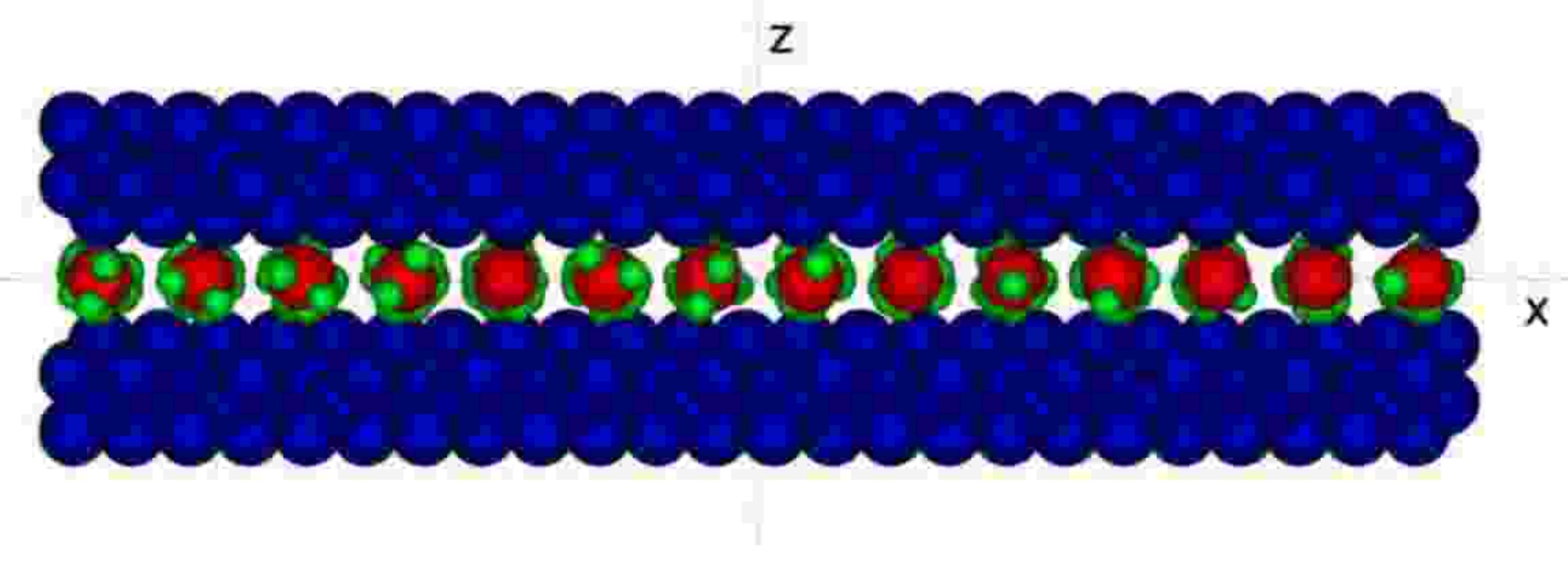}\includegraphics[width=0.49\textwidth]{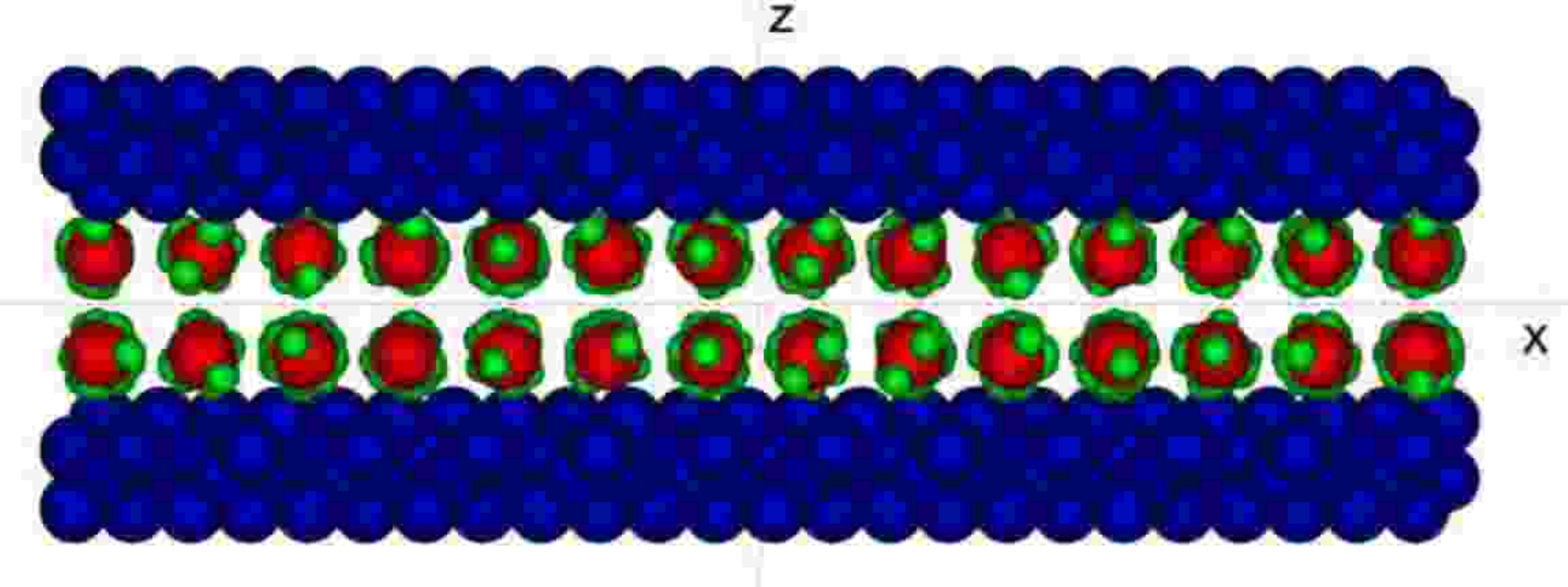}}
\caption{The initial configuration of the studied system with one (left) and two (right) layers of water molecules. Carbon, oxygen and hydrogen atoms are displayed as blue, red and green balls, respectively.}
\label{fig1}
\end{figure}

\begin{figure}[htb]
\centerline{\includegraphics[width=0.49\textwidth]{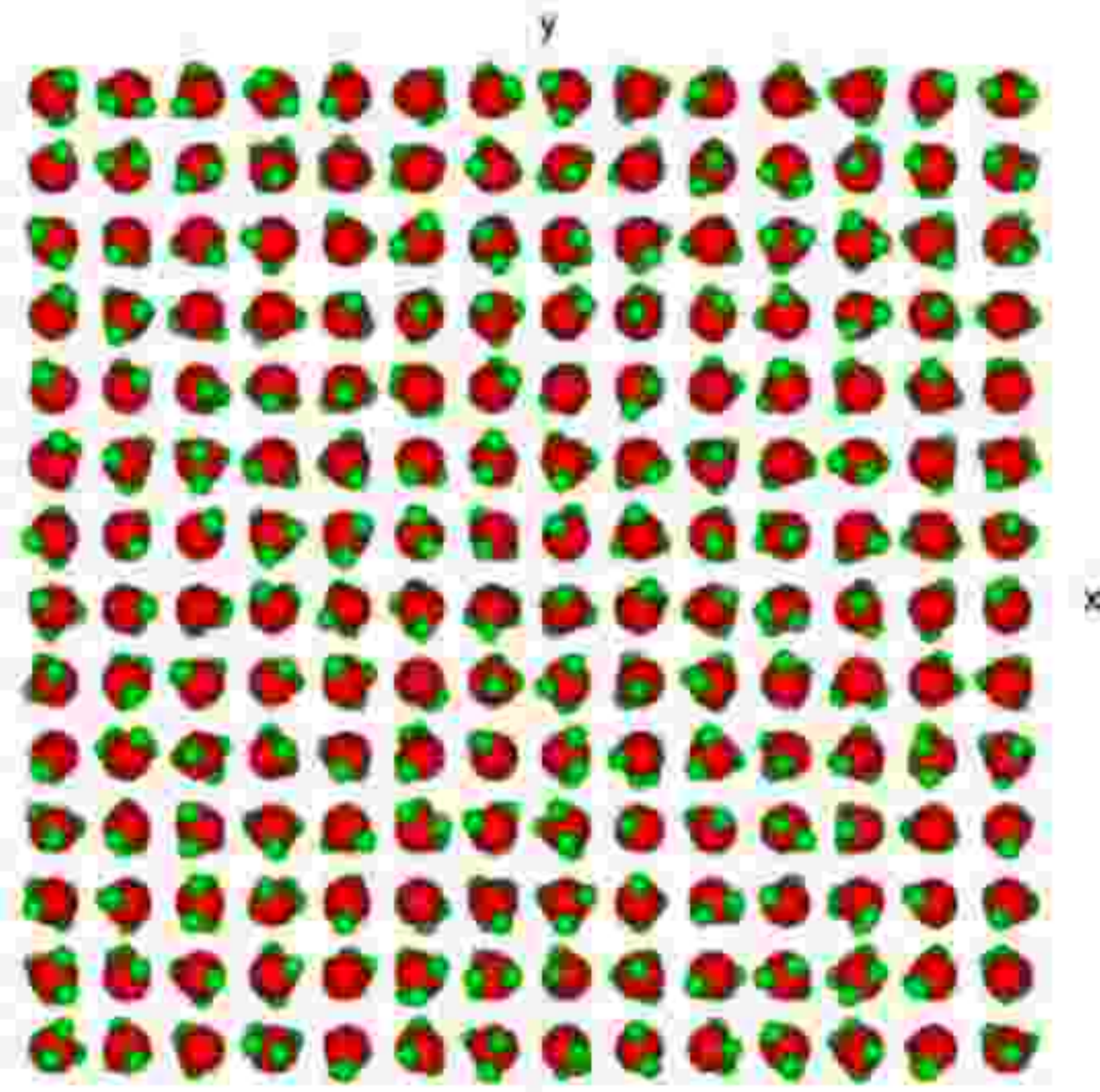}\includegraphics[width=0.49\textwidth]{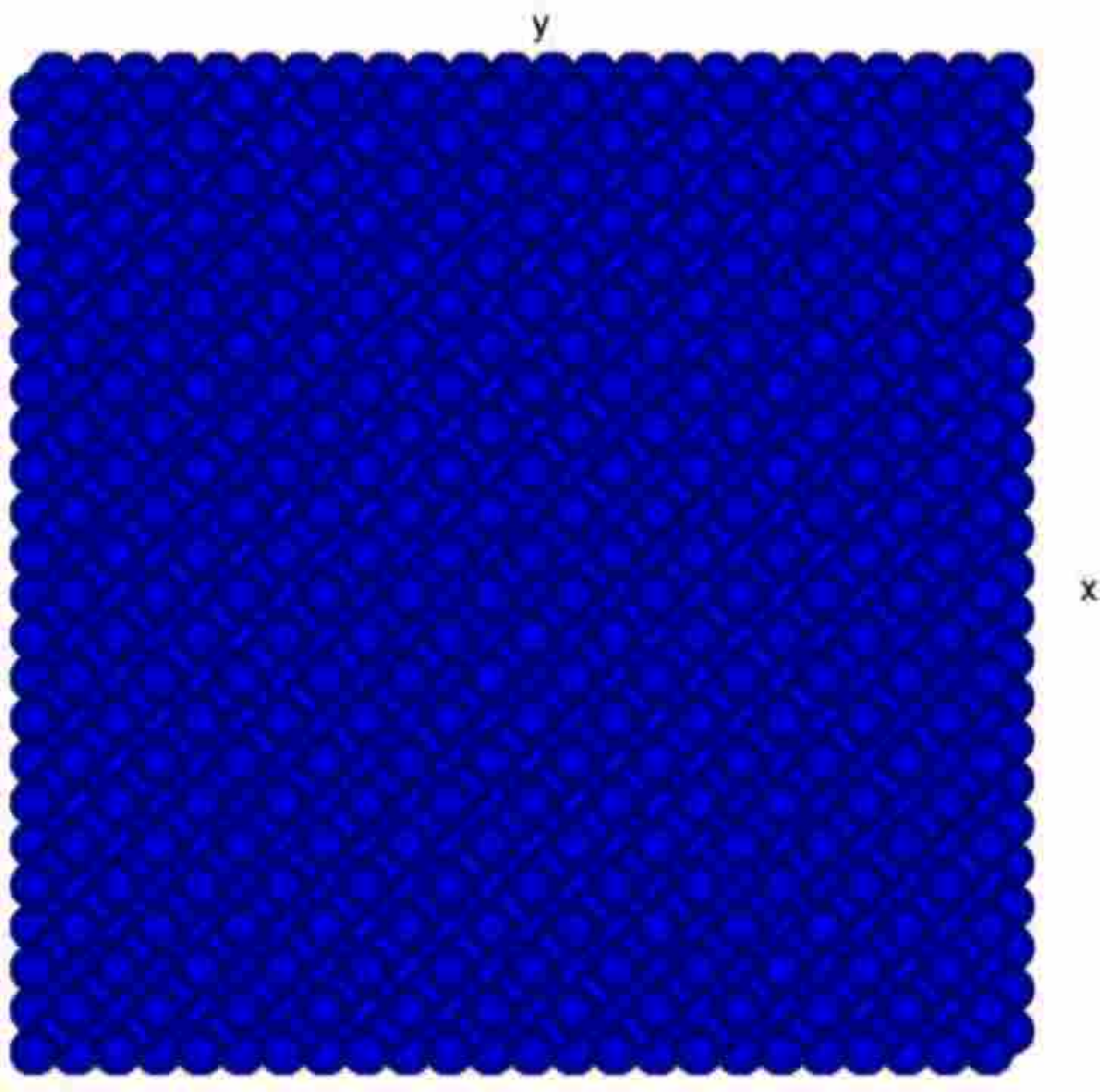}}
\caption{The arrangement of water molecules (left) and the form of diamond plate (right) in the \emph{xy} plane at the beginning of a simulation.}
\label{fig2}
\end{figure}

Figure~\ref{fig1} shows the initial configurations of the studied system for both values of the film's thickness. The initial state of the film and of the crystalline plate (view is from the top in the negative direction of the axis \emph{z}) is shown on figure~\ref{fig2}.

The empiric values of the covalent radii of atoms are used for visualization. For carbon, oxygen and hydrogen they are 77, 73, and 37~pm (1~pm = $10^{-12}$~m), respectively, the lattice constant of diamond is 356.68~pm \cite{webelemen}. At the beginning of simulations water molecules are positioned at the sites of a simple cubic lattice with the lattice constant corresponding to the density value of 1048~kg~m$^{-3}$. Molecular orientation is randomly assigned, with each linear and angular velocities having a fixed magnitude defined by the temperature and a randomly chosen directions. The water film and the plates are incommensurate because diamond lattice constant is not equal to the initial distance between water molecules. The initial distance (the gap) between the walls is 0.31~nm for one layer of water molecules and 0.62~nm for two layers. The horizontal dimensions of the computational cell are equal for \emph{x} and \emph{y} directions and their value is 42.81~\AA.

For imitation of the experiments the modelling is performed for constant temperature and load applied to the plates. The constraint method is used for maintaining the constant temperature value which is 298~K. The external load is modeled by applying the constant force $L$ to each atom of the walls along the \emph{z} axis. The constant-force algorithm is used \cite{Braun2001}. This means that shear is imitated by applying constant horizontal shearing or driving force $F_{\mathrm{S}}$ to each atom of the upper plate along the \emph{x} axis which corresponds to [010] direction.

The TIP4P model is used for water molecules \cite{Rapap2004}. It is based on four interaction sites located in planar configuration, two of which -- labeled M and O --  are associated with the oxygen nucleus, and two -- labeled H -- with protons. The interaction energy between two molecules $i$ and $j$ consists of a double sum over the interaction sites of both molecules; the terms in the sum, indexed by $k$ and $l$, allow for Coulomb interactions between the electric charges assigned to the sites, as well as contributions of LJ type,
\begin{equation}
\label{eq1-water-energy}
u_{ij}=\sum_{k\in i}\sum_{l\in j}\left(\frac{q_{k}q_{l}}{r_{kl}}+\frac{A_{kl}}{r_{kl}^{12}}-\frac{C_{kl}}{r_{kl}^{6}}\right),
\end{equation}
where $\textbf{r}_{kl}=\textbf{r}_{k}-\textbf{r}_{l}$ is the distance between sites \emph{k} and \emph{l}, $r_{kl}\equiv|\textbf{r}_{kl}|$. The corresponding force is
\begin{equation}
\label{eq2-water-force}
\textbf{f}_{ij}=\sum_{k\in i}\sum_{l\in j}\left(\frac{q_{k}q_{l}}{r_{kl}^{3}}+\frac{12A_{kl}}{r_{kl}^{14}}
-\frac{6C_{kl}}{r_{kl}^{8}}\right)\textbf{r}_{kl}.
\end{equation}
The charges appearing in the potential function are $q_{\mathrm{H}}=0.52e, q_{\mathrm{O}}=0, q_{\mathrm{M}}=-2q_{\mathrm{H}}$, where $e=1.6\times10^{-19}$~C. The parameters in the LJ part of the potential, which acts only between O sites, are $A_{\mathrm{O}\mathrm{O}}\equiv A=600\times10^{3}$~(kcal/mole)\AA$^{12}$, $C_{\mathrm{O}\mathrm{O}}\equiv C=610$~(kcal/mole)\AA$^{6}$, and 1~kcal/mole=4184~J/mole.

The dimensionless units are used in the simulations. The length unit is $\sigma=\left(A/C\right)^{1/6}=3.154$~\AA, the energy unit is $\varepsilon=A/\left(4\sigma\right)^{12}=0.155$~kcal/mole, the mass unit is the mass of a water molecule, $m=2.987\times10^{-27}$~g and the time unit is $t_{0}=\sqrt{m\sigma^{2}/\varepsilon}=5.253\times10^{-12}$~s. Also in dimensionless units $q_{\mathrm{H}}=1$. The time step used is $\Delta t=0.0005$; in real units this equals to $2.627\times10^{-15}$~s.

For the water-diamond interactions only the interactions of O site of the water molecule with carbon atoms are assumed. The interaction potential is of LJ type
\begin{equation}
\label{eq3-water-diamond}
u_{ij}=\left\{\begin{array}{lr}
         4\varepsilon_{\mathrm{C}\mathrm{O}}\left[
         \left(\frac{\sigma_{\mathrm{C}\mathrm{O}}}{r_{ij}}\right)^{12}-
         \left(\frac{\sigma_{\mathrm{C}\mathrm{O}}}{r_{ij}}\right)^{6}\right], & r_{ij} < r_{\mathrm{c}}\\
         0, & r_{ij} \geq r_{\mathrm{c}}
       \end{array}\right.,
\end{equation}
where $r_{\mathrm{c}}=7.5$~\AA (or 2.38 in dimensionless units) is the interaction cutoff distance, and $\varepsilon_{\mathrm{C}\mathrm{O}}=2$, $\sigma_{\mathrm{C}\mathrm{O}}=0.86$. The values of latter parameters are chosen to fit the experimental fact that diamond is hydrophilic \cite{Vojut1975}. They correspond to the situation when the attractive forces between water molecules and carbon atoms have a twice greater magnitude than intermolecular water forces. The interaction between atoms of different plates is not introduced into the model.

The classical equations of motion are used in simulations. As a consequence of the rigidity of diamond plates the motion of their centers of masses is concerned. The equations of motion for the upper plate are
\begin{equation}
\label{eq4-motion-uppl}
\begin{array}{c}
  M\ddot{X}=F_{\mathrm{x}}+F_{\mathrm{S}}N_{\mathrm{p}},\\
  M\ddot{Y}=F_{\mathrm{y}}, \\
  M\ddot{Z}=F_{\mathrm{z}}+LN_{\mathrm{p}},
\end{array}
\end{equation}
where $X,Y,Z$ are the coordinates of the center of mass of the upper plate, $N_{\mathrm{p}}$ is the number of atoms of the wall, $M=N_{\mathrm{p}}m_{\mathrm{C}}$ is the plate's mass, $m_{\mathrm{C}}=0.67$ is the carbon atomic mass, $F_{\mathrm{x}},F_{\mathrm{y}},F_{\mathrm{z}}$ are the components of the total force acting on the wall from the water. They are defined in a standard manner as the sum of the appropriate coordinate derivatives of the potential (\ref{eq3-water-diamond}) with the negative sign for all molecules. Equations of motion for the lower plate are similar to (\ref{eq4-motion-uppl}). The difference is in that there is no component responsible for shear as there is in the first equation in (\ref{eq4-motion-uppl}).

The equations of the translational motion (without a component responsible for the constancy of temperature) for the center of mass of the $i$th molecule are as follow:
\begin{equation}
\label{eq5-trmotion-water}
\ddot{\textbf{r}}_{i}=\sum_{j}\textbf{F}_{ij}+\sum_{k}\textbf{f}_{ik},
\end{equation}
where $\sum_{j}\textbf{F}_{ij}$ is the force acting from all carbon atoms that are not further that $r_{\mathrm{c}}$ from the $i$th atom and it is defined in the way similar to the force for the surfaces; $\sum_{k}\textbf{f}_{ik}$ is the force acting on the $i$th molecule from other water molecules inside the cutoff distance, it is defined by the equation (\ref{eq2-water-force}). The rotational motion of water molecules is taken into account in the model and the equations of the rotational motion are expressed in terms of Hamilton's quaternions \cite{Rapap2004}.

\section{Measurements, results and discussion}

Molecular dynamics simulations are performed for values of the load $L$ per atom in the range from 2 (which is equal to 6.838~pN in real units) to 50 (0.171~nN). Corresponding pressure acting on the plates varies from 0.43 to 10.751~GPa. The horizontal shearing force $F_{\mathrm{S}}$ acting on each atom of the upper plate has values from 0.5 (or 1.71~pN) to 10 (34.3~pN) per atom and the resulting shear force acting on the plate changes from 1.97 to 394~nN. The maximum duration of the simulation is 52000 time steps or 136.6~ps. At each simulation during the first 2000 time steps the system is equilibrated and after that the measurements are carried out. There are two groups of simulations. The first group is characterized by the absence of shearing force and the measurement of the diffusion constant is performed. In the second group the shear of the upper plate takes place and the friction force is measured. The latter is presented by the first component in (\ref{eq4-motion-uppl}).

The diffusion constant $D$ is calculated in two ways. In the first case the Einstein expression is used \cite{Rapap2004}
\begin{equation}
\label{einstein}
D=\lim_{t\rightarrow\infty}\frac{1}{6N_{\mathrm{m}}t}\langle\sum_{j=1}^{N_{\mathrm{m}}}
\left[\textbf{r}_{j}(t)-\textbf{r}_{j}(0)\right]^{2}\rangle,
\end{equation}
where $N_{\mathrm{m}}$ is the number of molecules, $t$ is the time of measurements and $\textbf{r}_{j}$ is the radius vector of the $j$th molecule center of mass. In (\ref{einstein}) angled brackets denote the average over a sufficiently large number of independent samples.

The second way for calculating the diffusion constant is using the alternative Green-Kubo expression for $D$ based on the integrated velocity autocorrelation function $\varphi(t)$ \cite{Rapap2004,March1976}
\begin{equation}
\label{Green-Kubo}
D=\frac{1}{3N_{\mathrm{m}}}\int_{0}^{\infty}\varphi(t)\mathrm{d}t,
\end{equation}

\begin{equation}
\label{acfv}
\varphi(t)=\langle\sum_{j=1}^{N_{\mathrm{m}}}\textbf{\emph{v}}_{j}(t)\cdot\textbf{\emph{v}}_{j}(0)\rangle,
\end{equation}
where $\textbf{\emph{v}}_{j}$ is the velocity of the $j$th molecule.

\begin{figure}[htb]
\centerline{\includegraphics[width=0.5\textwidth]{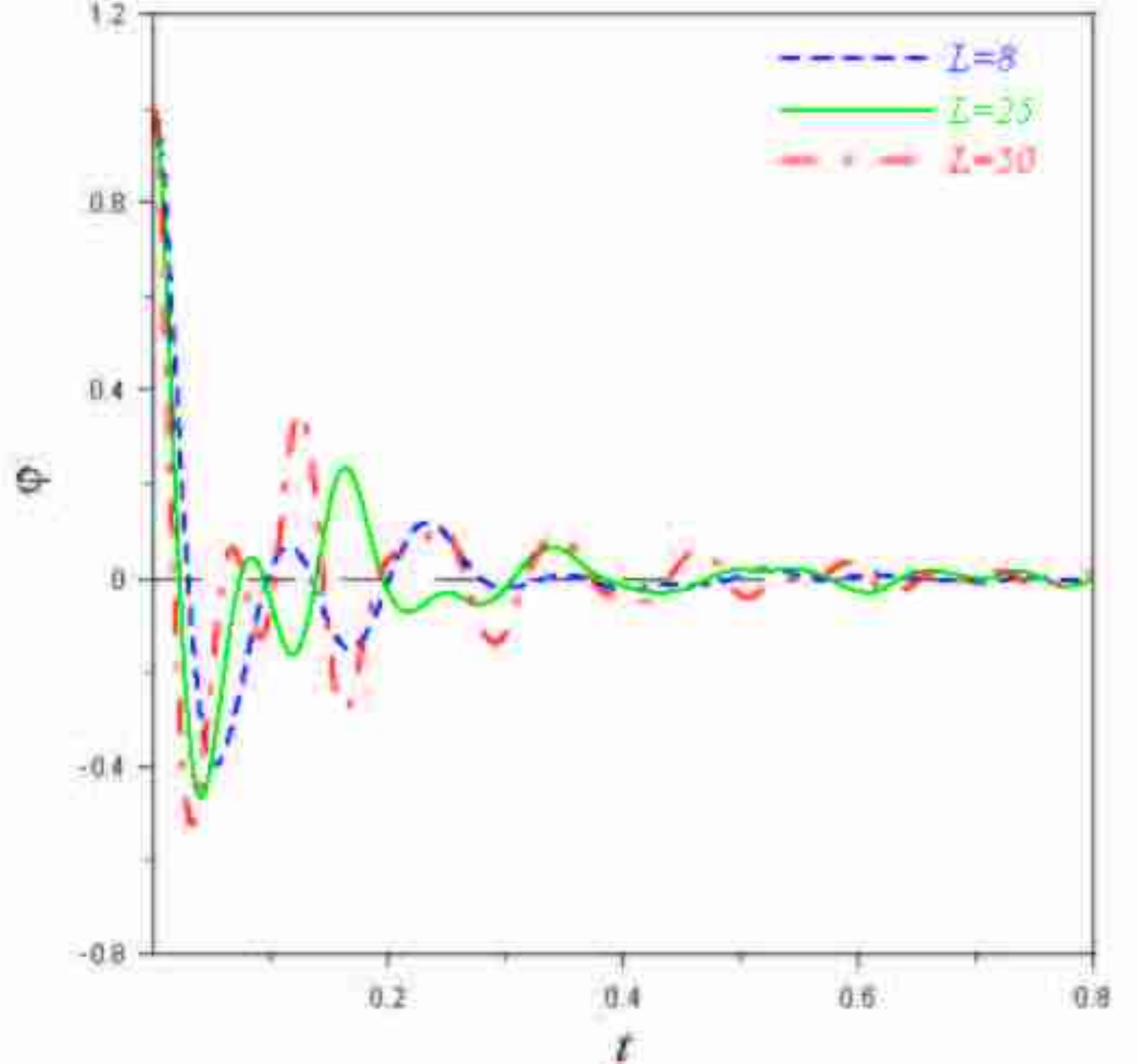}\includegraphics[width=0.5\textwidth]{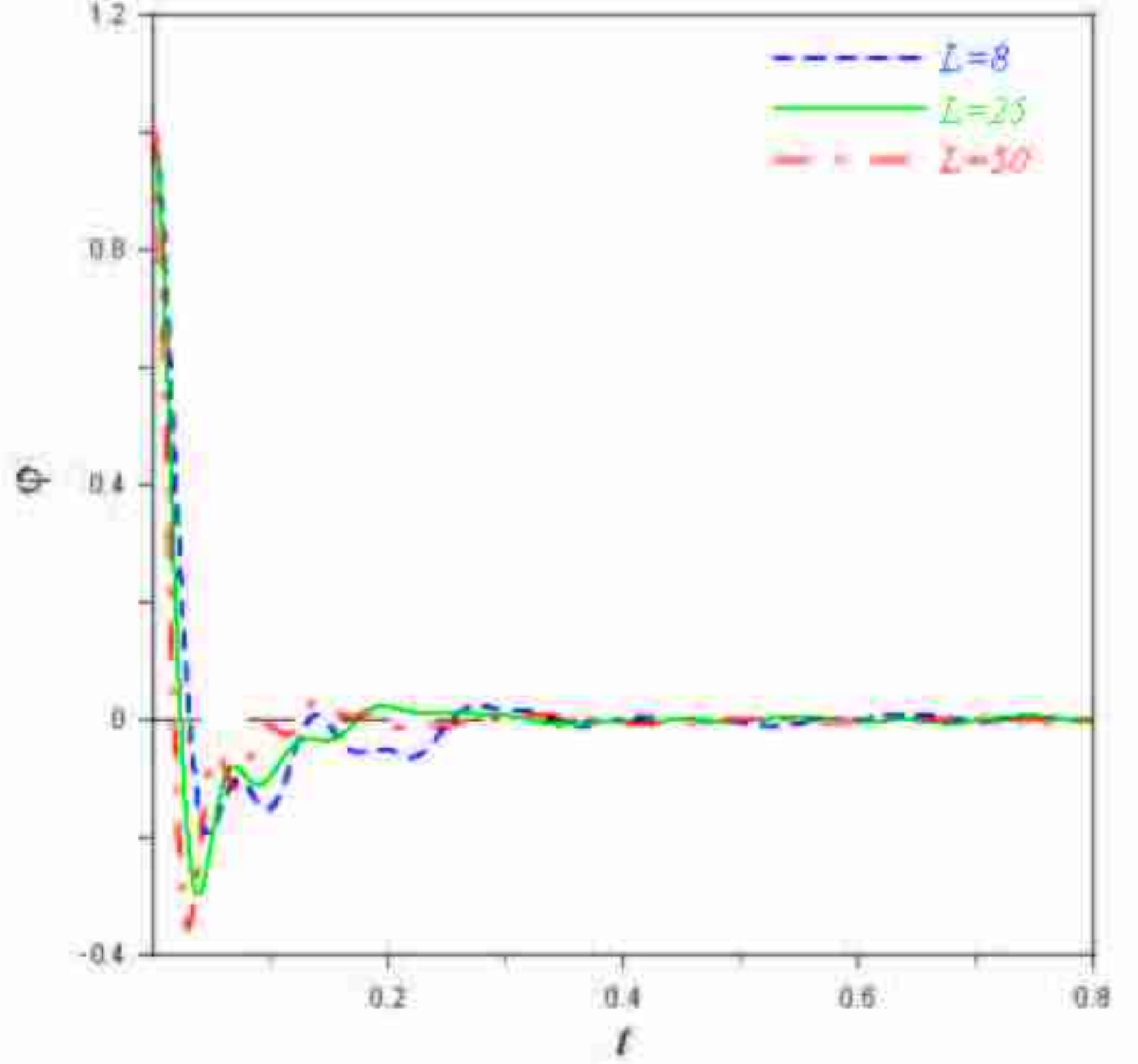}}
\caption{Time dependencies of the velocity autocorrelation function (\ref{acfv}) for different loads and films with thickness of one (left) and two (right) molecular diameters.}
\label{fig3}
\end{figure}

Figure~\ref{fig3} shows calculated time dependencies of the velocity autocorrelation function of water molecules for different loads. It is apparent that with the increase of load the number and frequency of oscillations increases. This indicates that molecule velocities become more correlated under higher loads.

\begin{figure}[h]
\centerline{\includegraphics[width=0.45\textwidth]{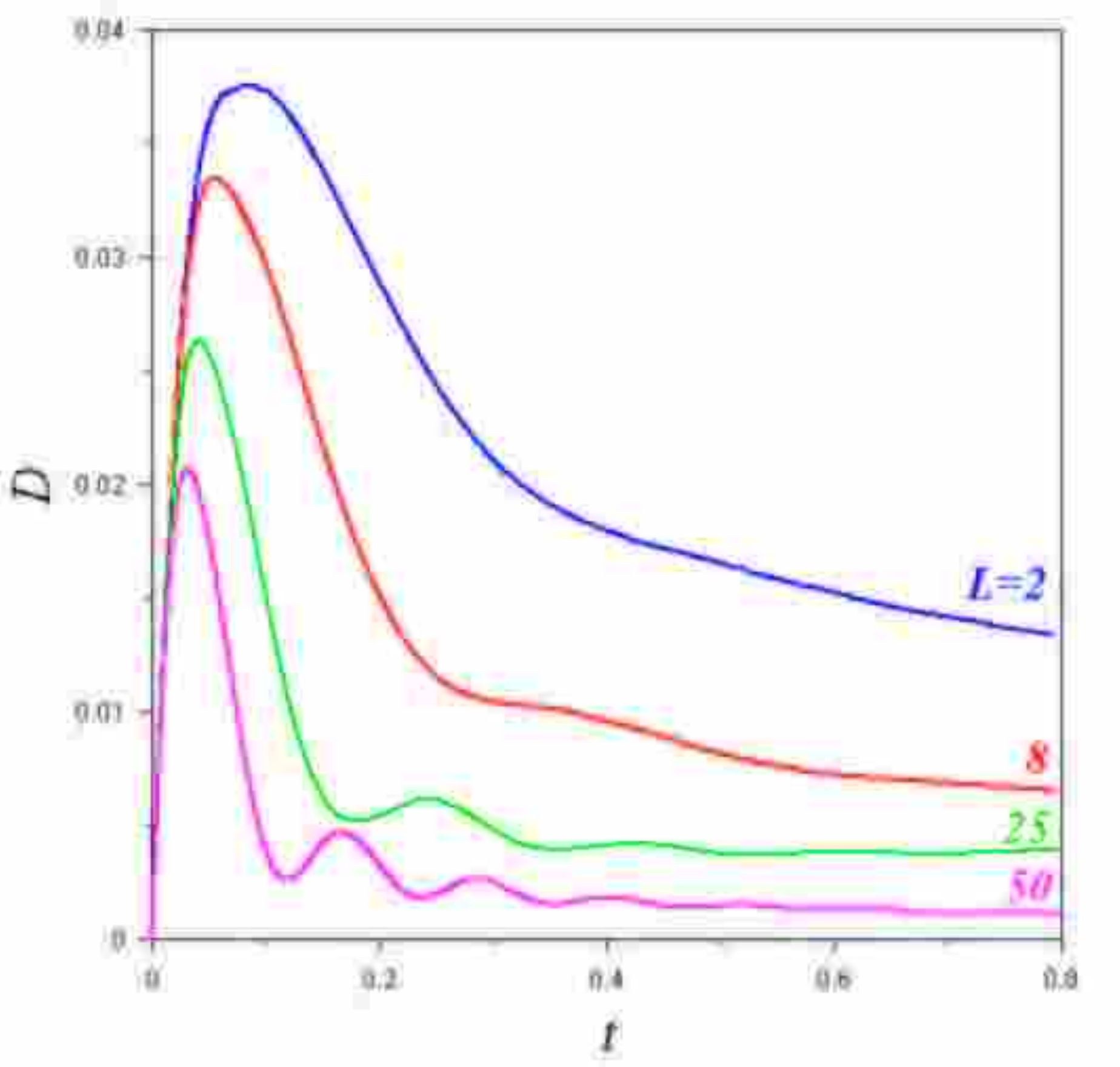}\includegraphics[width=0.45\textwidth]{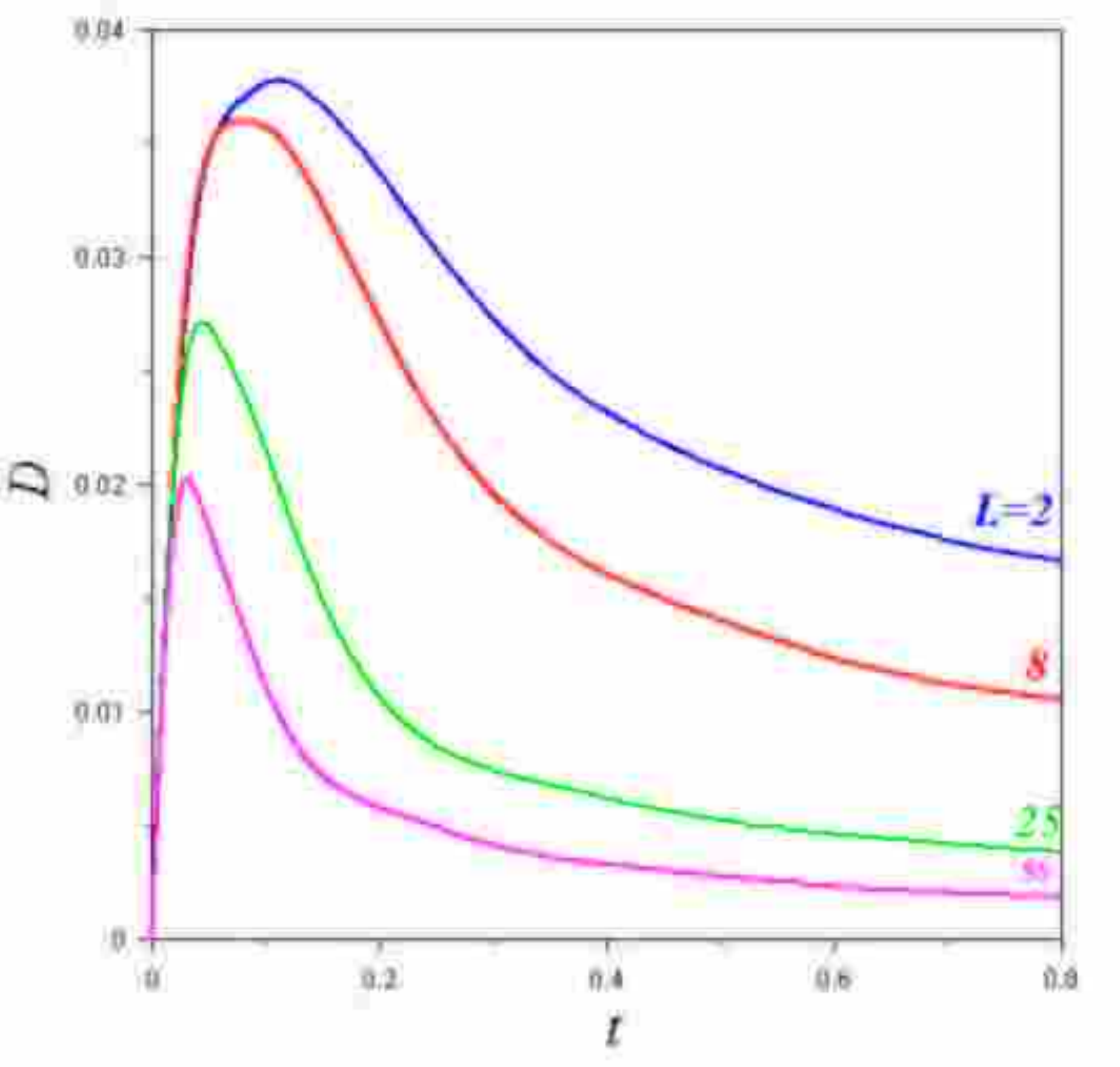}}
\caption{Time dependencies of the diffusion constant for different loads and films with thickness of one (left) and two (right) molecular diameters. Values are calculated with the Einstein expression~(\ref{einstein}).}
\label{fig4}
\end{figure}

\begin{figure}[!]
\centerline{\includegraphics[width=0.45\textwidth]{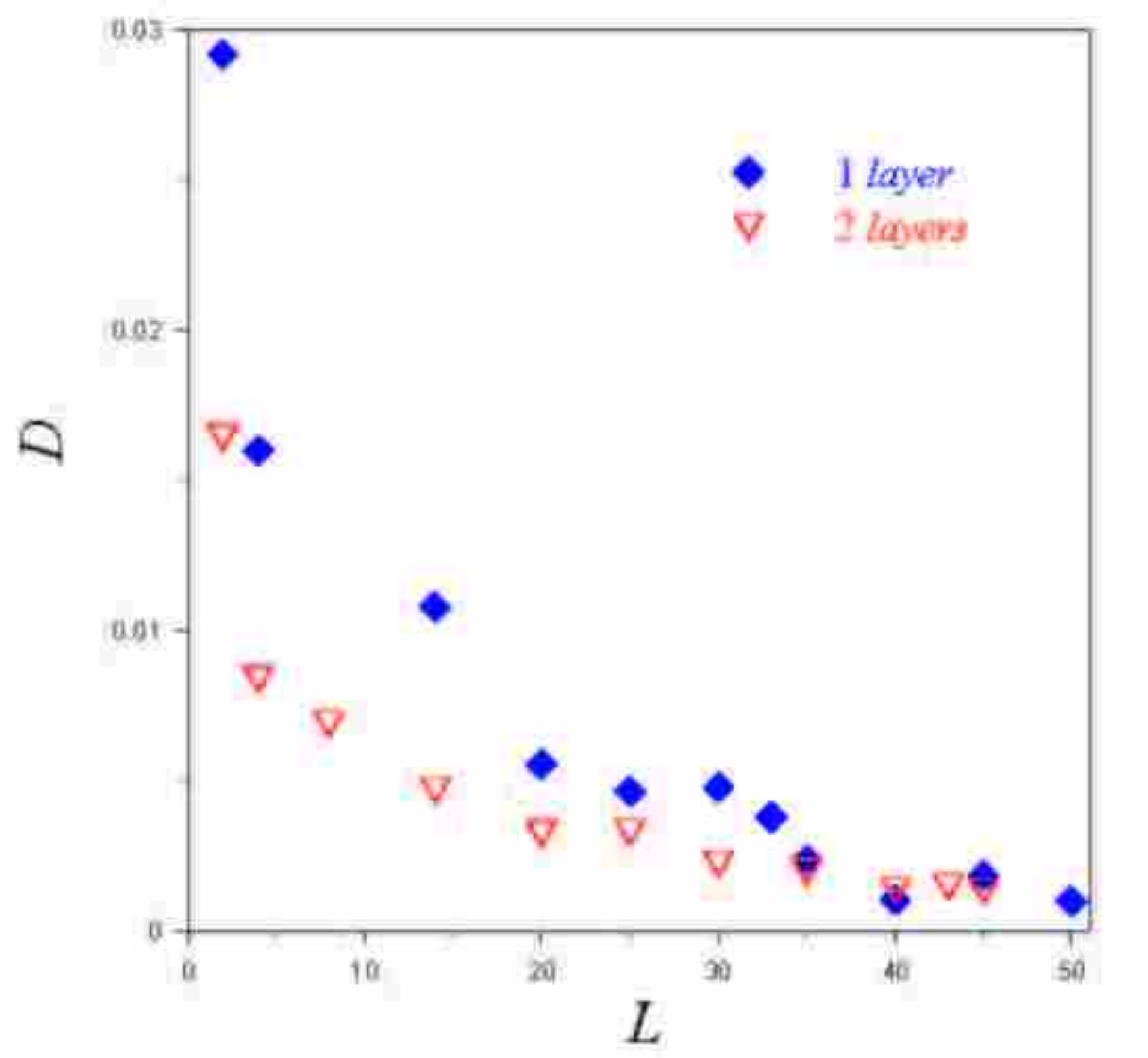}}
\caption{The diffusion constant calculated with the Green-Kubo formula~(\ref{Green-Kubo}) as function of load for films with thickness of one and two molecular diameters.}
\label{fig5}
\end{figure}

\begin{figure}[h]
\centerline{\includegraphics[width=0.45\textwidth]{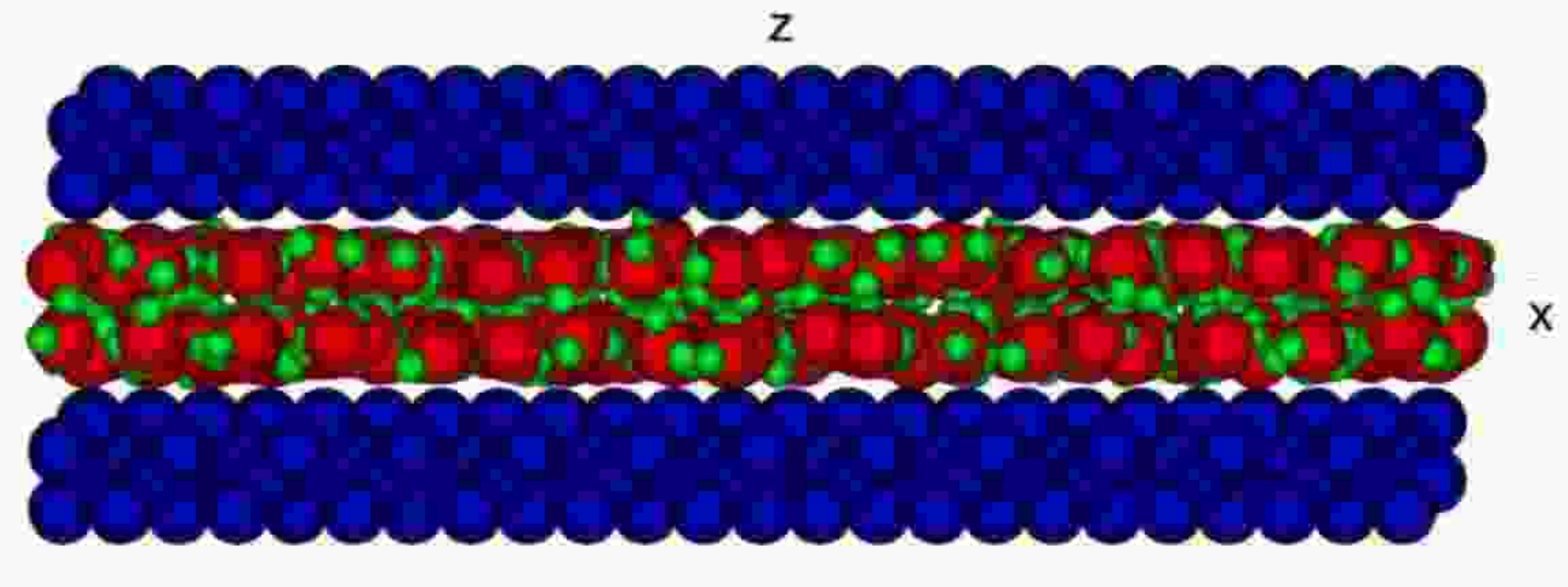}\includegraphics[width=0.45\textwidth]{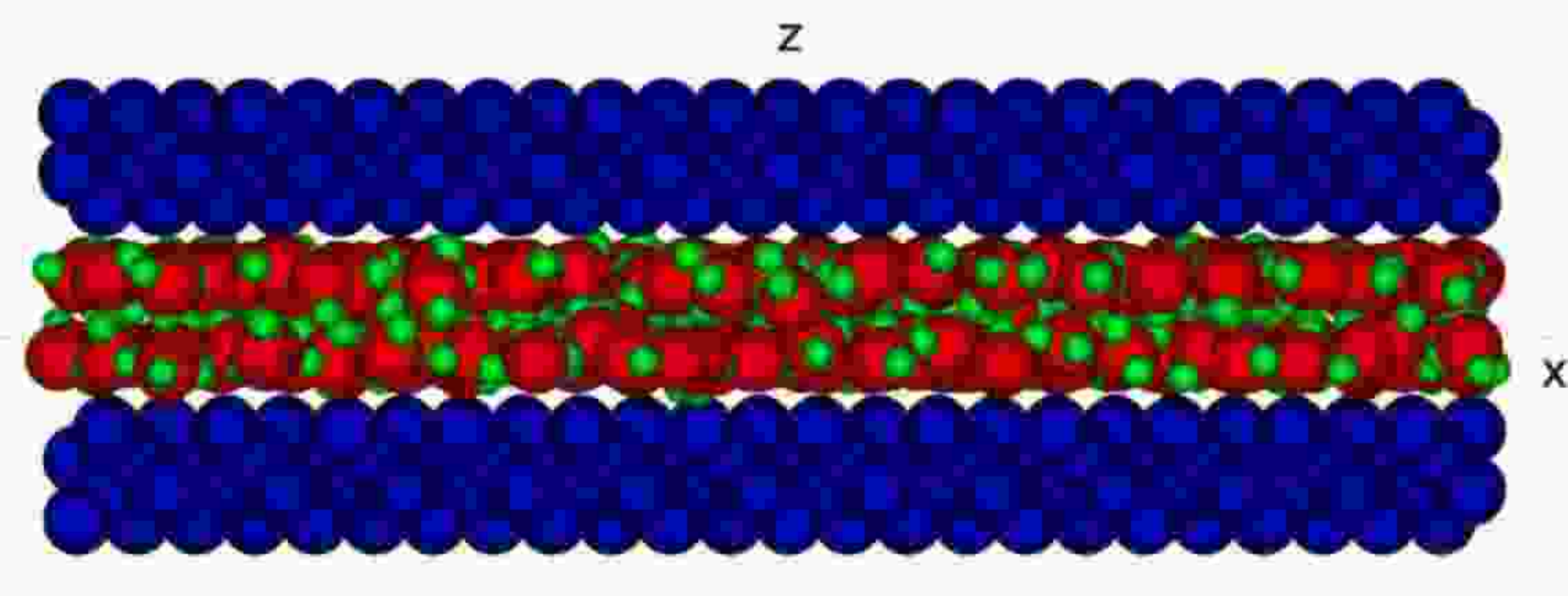}}
\caption{Layering of water molecules for values of load 8 (left) and 14 (right) and zero shear force.}
\label{fig6}
\end{figure}

\begin{figure}[!t]
\centerline{\includegraphics[width=0.5\textwidth]{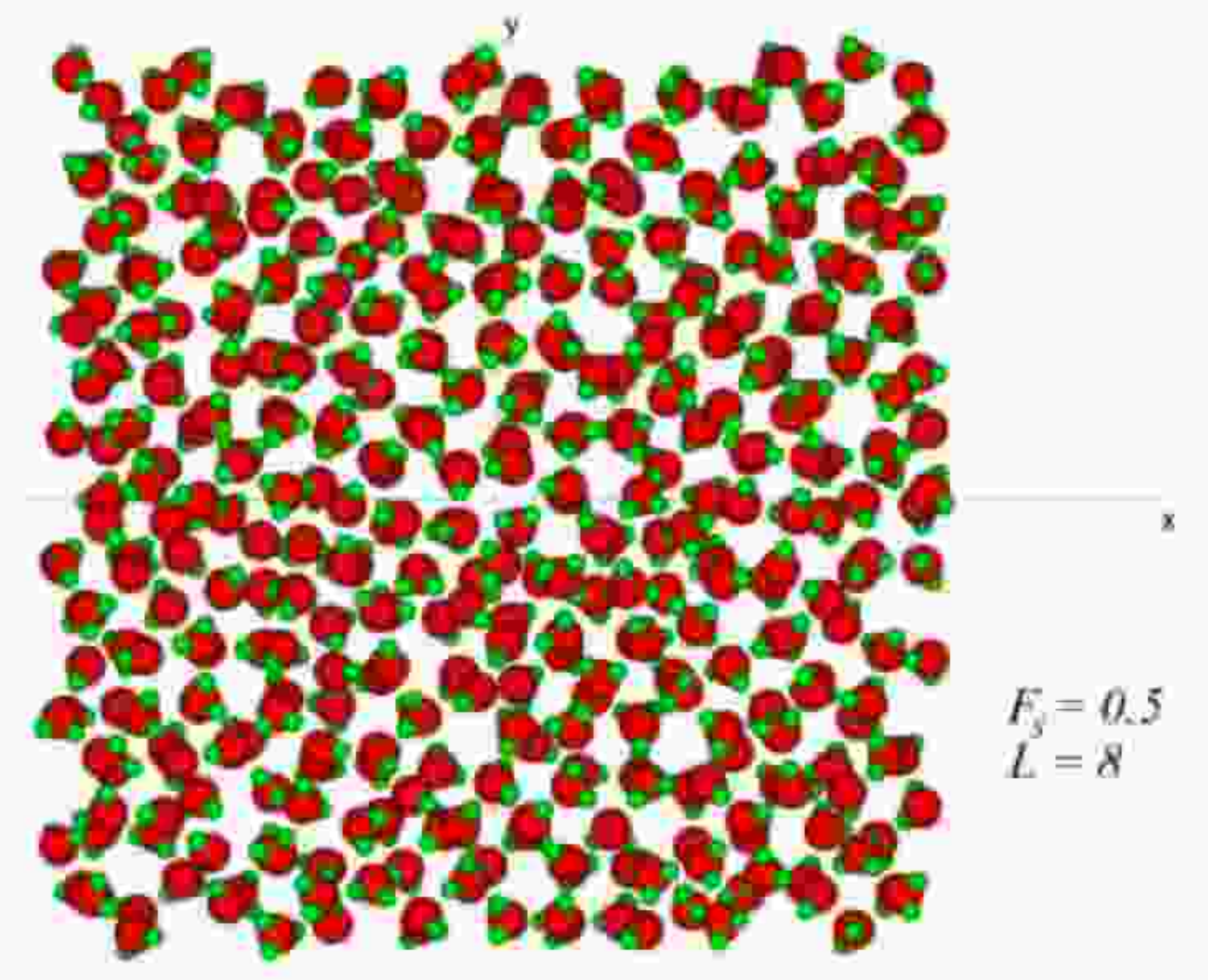}\includegraphics[width=0.5\textwidth]{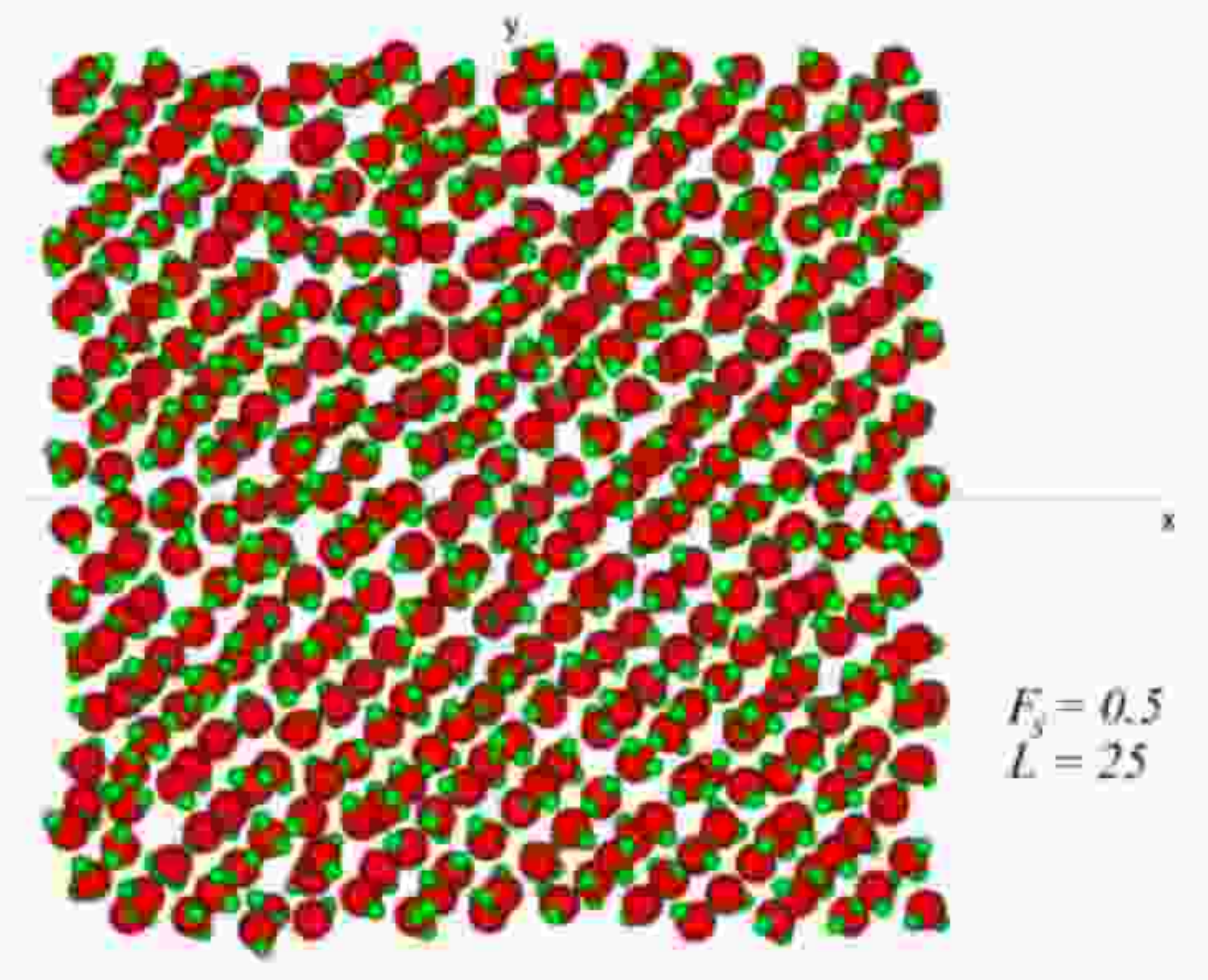}}
\centerline{\includegraphics[width=0.5\textwidth]{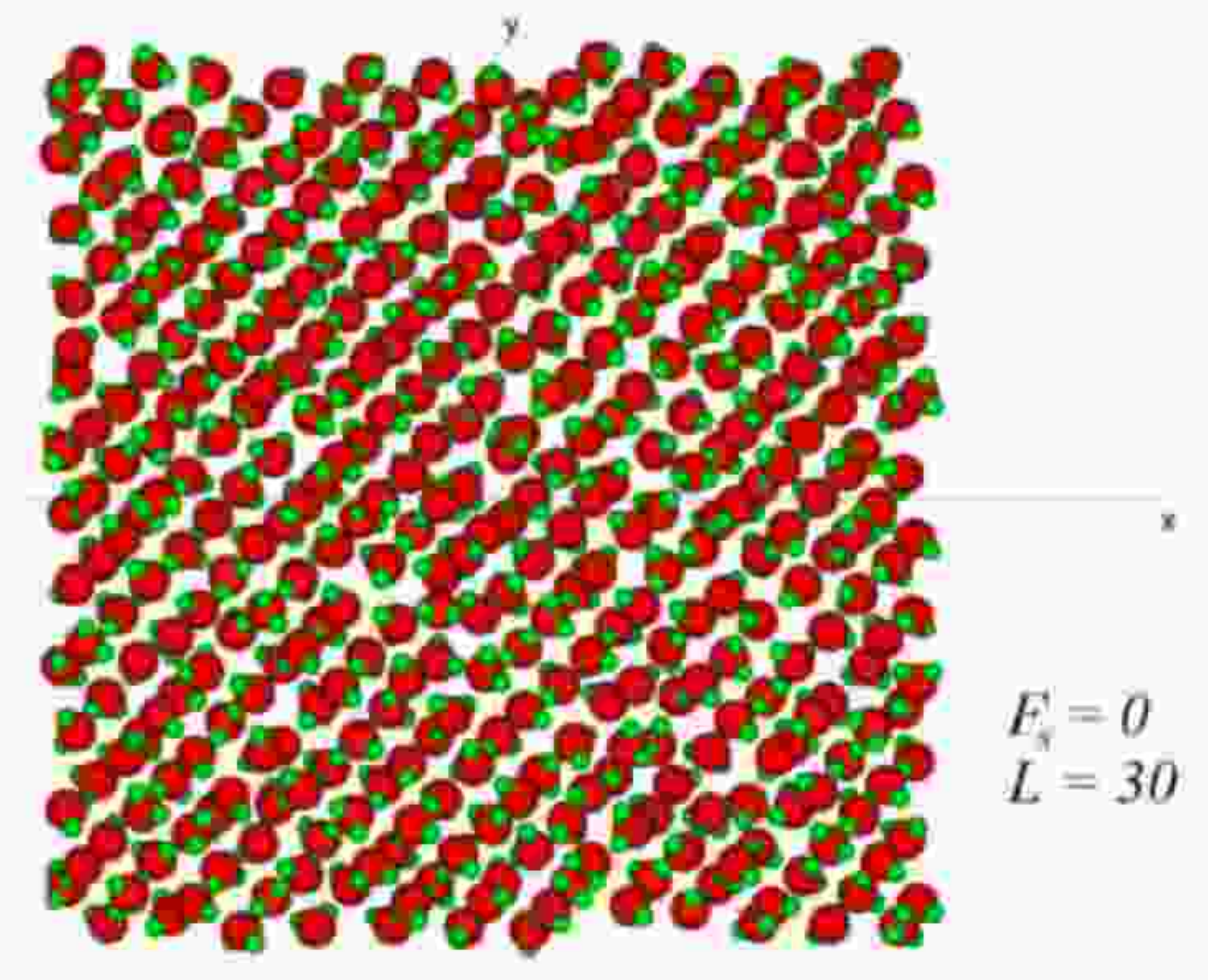}\includegraphics[width=0.5\textwidth]{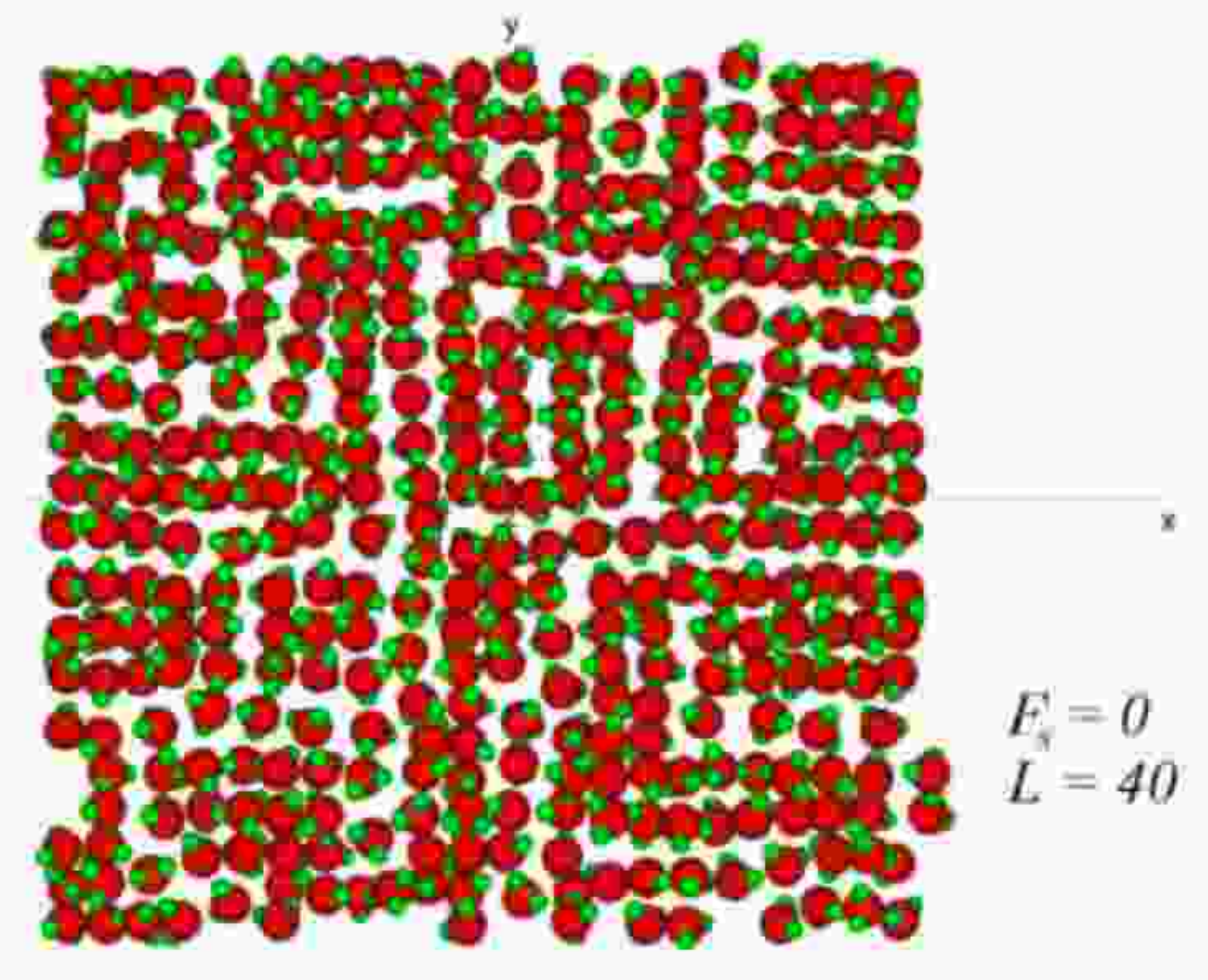}}
\centerline{\includegraphics[width=0.5\textwidth]{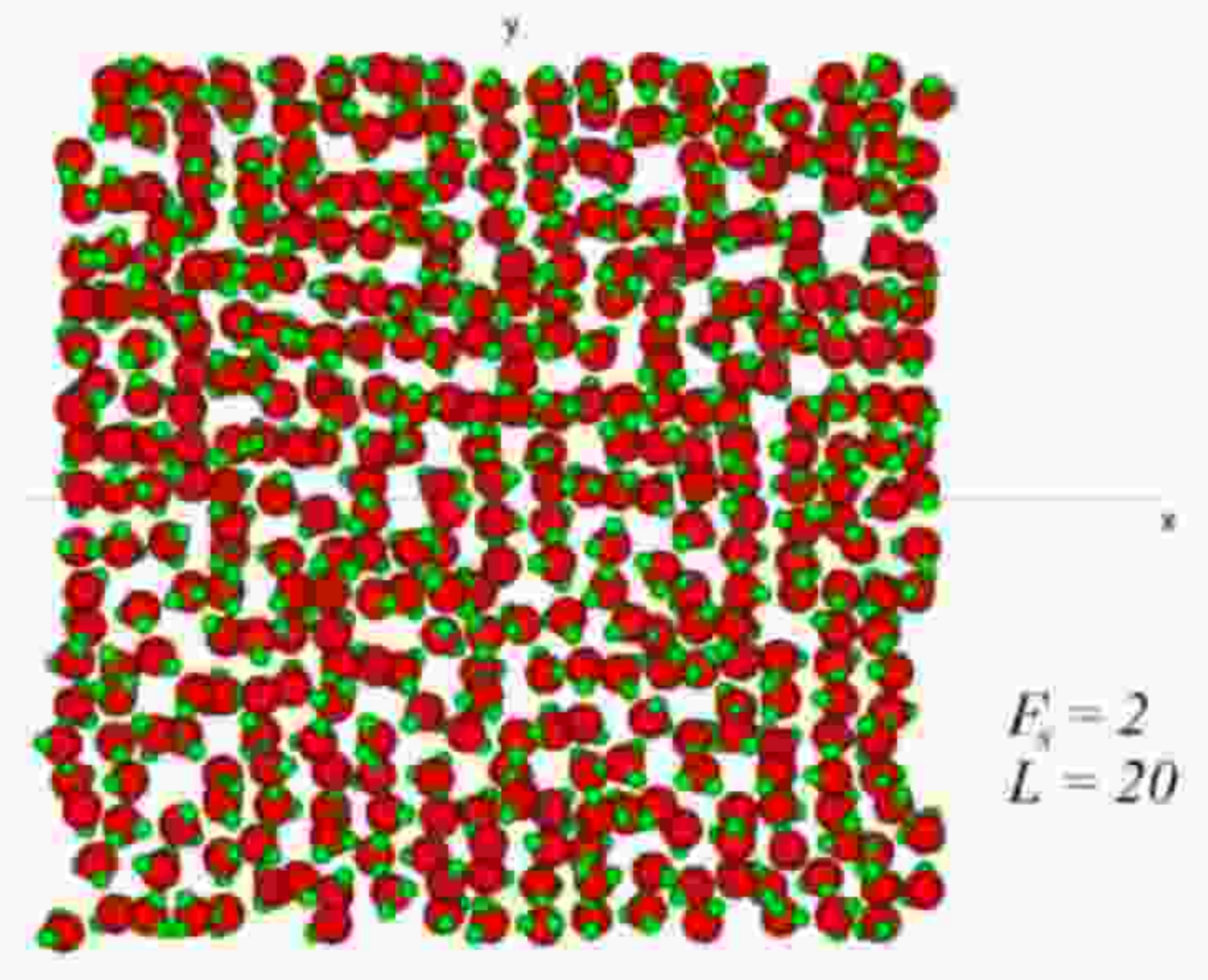}\includegraphics[width=0.5\textwidth]{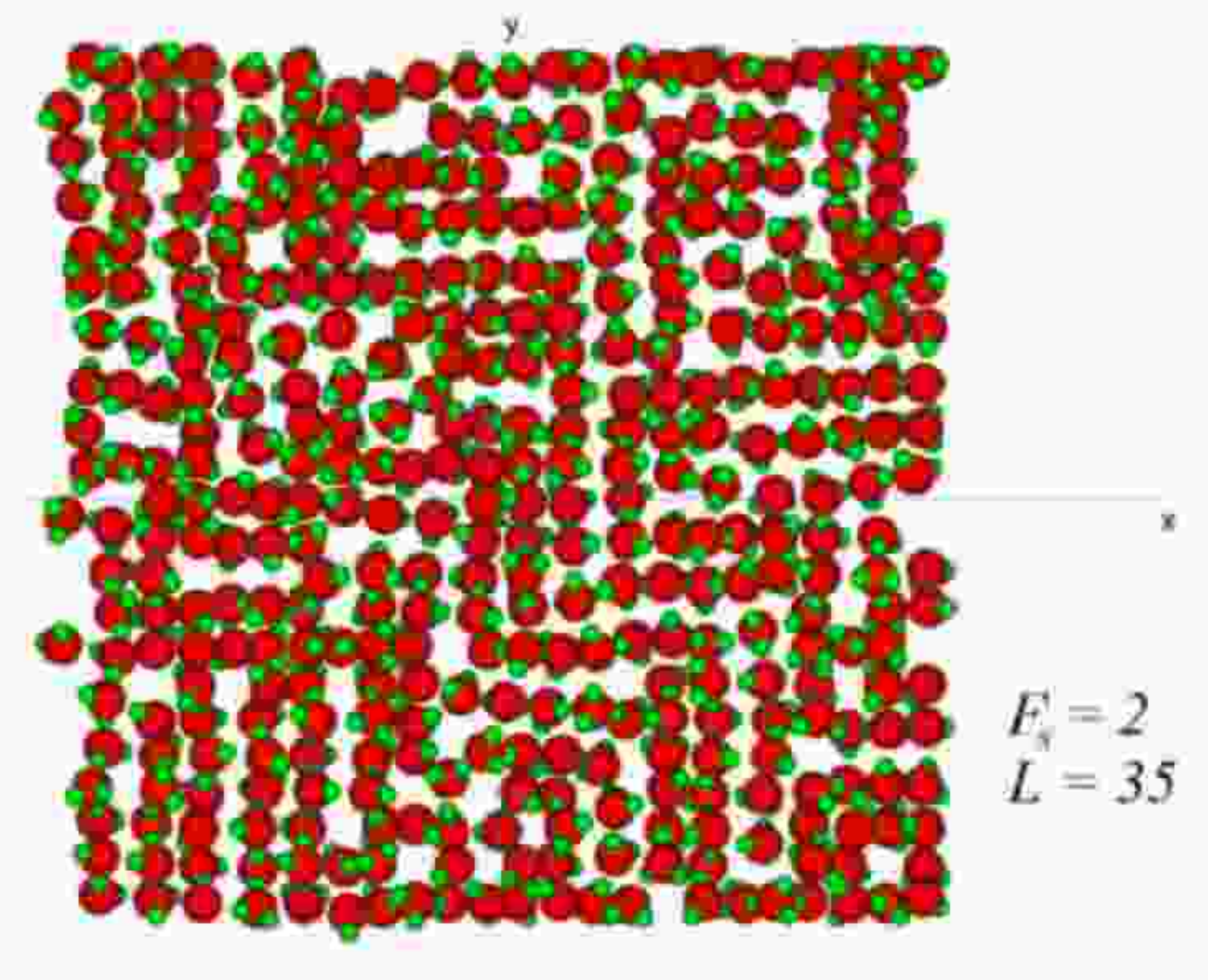}}
\caption{Configurations of molecules in \emph{xy} plane  for different loads and shear forces at the end of corresponding simulations with the two-layer water film.}
\label{fig7}
\end{figure}

A gradual decrease of the diffusion constant with the external load increase shown on figure~\ref{fig4} and figure~\ref{fig5} indicates the transition of the water film to a solidlike state. The feature of this state is the ordering of molecules which can be vertical (in the form of quasidiscrete layers) and horizontal (or in-plane). Analytically the former manifests itself in the density oscillations with distance normal to the boundaries and the measure of the latter is the Debye-Waller factor. In this work mentioned quantities have not been calculated but ordering is observed visually. Figure~\ref{fig6} shows typical vertical ordering of water molecules for two values of load. In-plain ordering for one-layer film is not observed for all used values of load and driving force. However, for two molecular diameters thick film in-plane ordering takes place especially for high loads. Typical molecule configurations are shown on figure~\ref{fig7}. Here one can see disordered and several types of ordered configurations of molecules. The presence of shear, as a rule, promotes the occurrence of the ordered structure for lower loads than without shearing. Also shear results in more rapid ordering. Thus, for two-layer thick water film there is a shear ordering that is observed experimentally \cite{Yoshi1993}. It should be noted that for low driving forces and intermediate loads molecular configuration is not completely defined by parameters of these forces. For example, structures shown on figure~\ref{fig7} for $F_{\mathrm{S}}=0.5, L=25$ and $F_{\mathrm{S}}=0, L=30$ are not absolutely reproducible and in other simulations for these values disordered configurations are obtained. In contrast to such behaviour for higher loads ordered state is always obtained as shown on the bottom right part of figure~\ref{fig7}.

The absence of ordering in the one-layer film we can explain after realizing the model peculiarities. The behaviour of one-layer film should be similar to the one of a simple LJ liquid confined between unstructured (i.e. mathematically smooth) surface. This follows from the facts that the roughness of surfaces is negligible and in such thin water film in the $\emph{z}$ direction its molecules can interact only with surface atoms. Surface atoms in turn interact only with O sites of water molecules and only through the LJ potential. Therefore LJ contribution dominates in the interaction of water molecules in one-layer film. For the thicker film water molecules in adjacent layers can interact with each other and Coulomb interaction can cause the occurrence of ordered structure. It should be noted that observed ordering is not the true thermodynamic phase transition. And a liquidlike and a solidlike states are not really the same as the bulk liquid or solid phases. It would be more correct to refer to them perhaps as static and dynamic ``epistates'', since they arise only in boundary films, whose properties are determined both by the confinement and the epitaxial interactions between the film and surface atoms \cite{Yoshi1993}.

\begin{figure}[h]
\centerline{\includegraphics[width=0.48\textwidth]{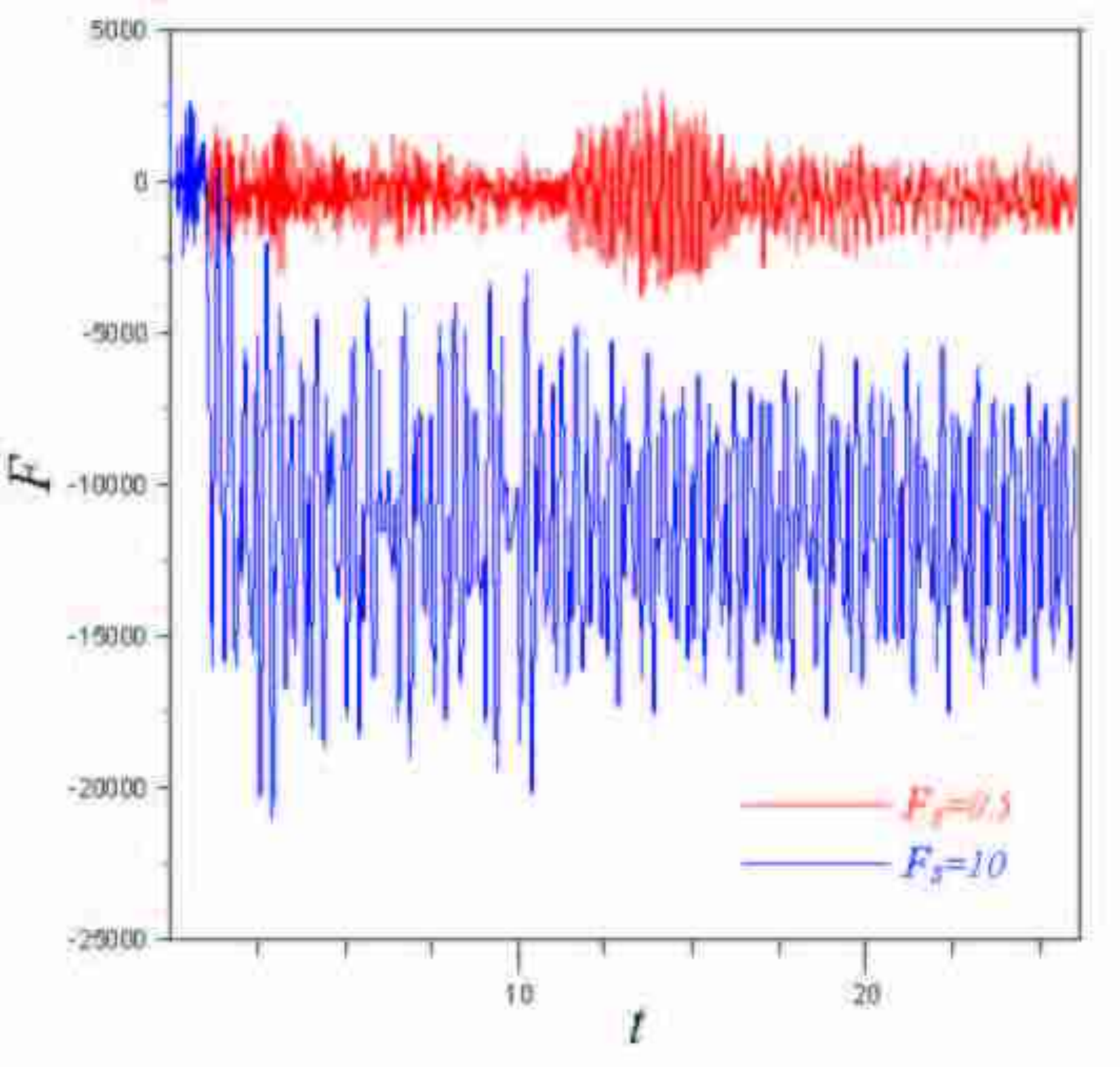}\includegraphics[width=0.48\textwidth]{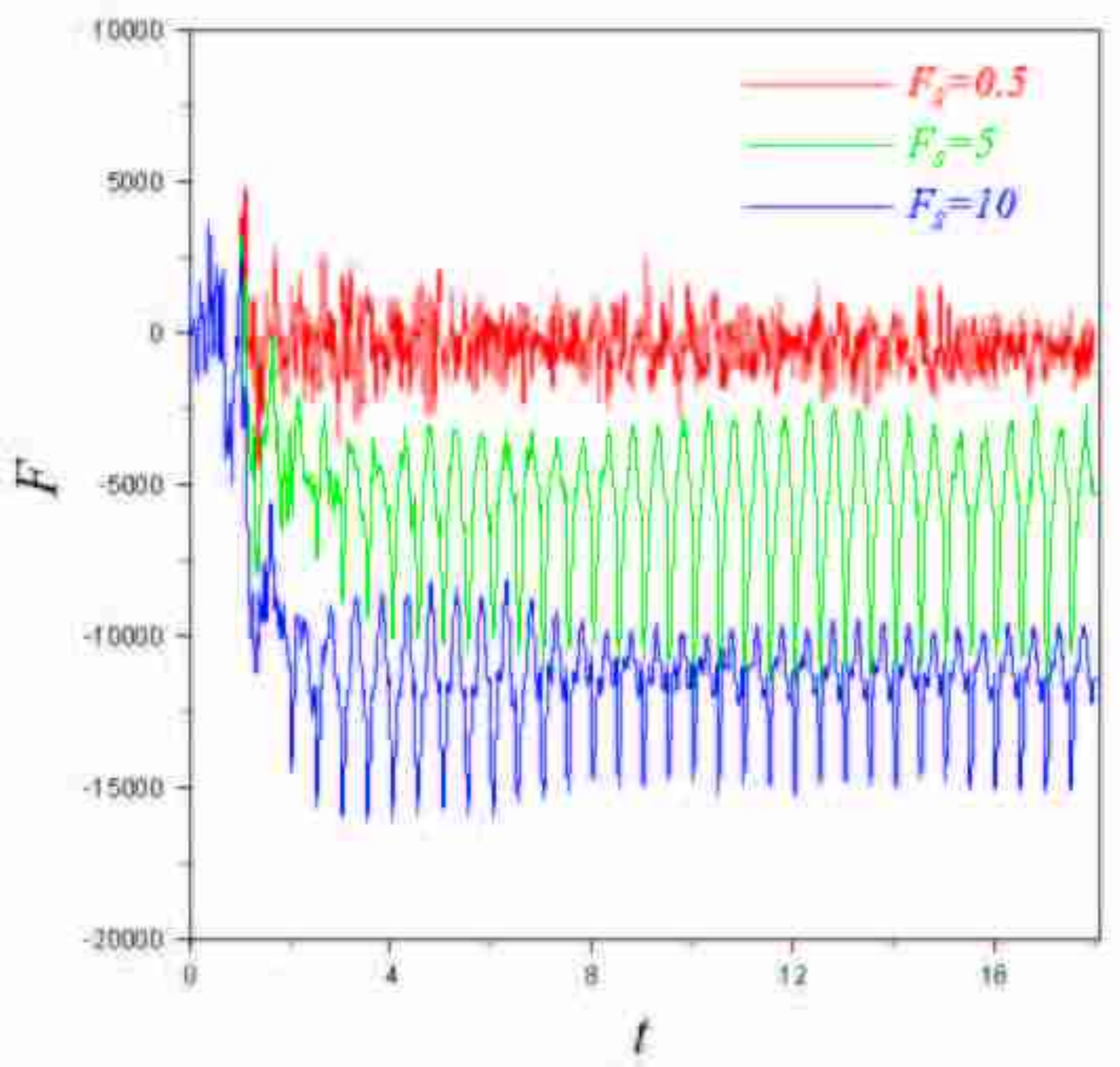}}
\caption{Time dependencies of the friction force for films with thickness of one (left) and two (right) molecular diameters and different driving forces. The value of load is 50. Negative trend is the consequence of the fact that friction force acts in the direction reverse to sliding.}
\label{fig8}
\end{figure}

Figure~\ref{fig8} shows time dependencies of friction force obtained for different values of shearing force. For one-layer film such dependency is irregular with considerable fluctuations. For the thicker film periodic spikes are observed for higher values of shearing force. Similar spikes are observed in experiments when the film is in a solidlike state and they correspond to the stick-slip friction \cite{Yoshi1993,GeeML1990}. But the considerable difference in time values in simulations and experiments which is 10 orders of magnitude prevents the definite comparison of the results of simulations with experiments (in experiments time values are of order of seconds-minutes).

\begin{figure}[h]
\centerline{\includegraphics[width=0.48\textwidth]{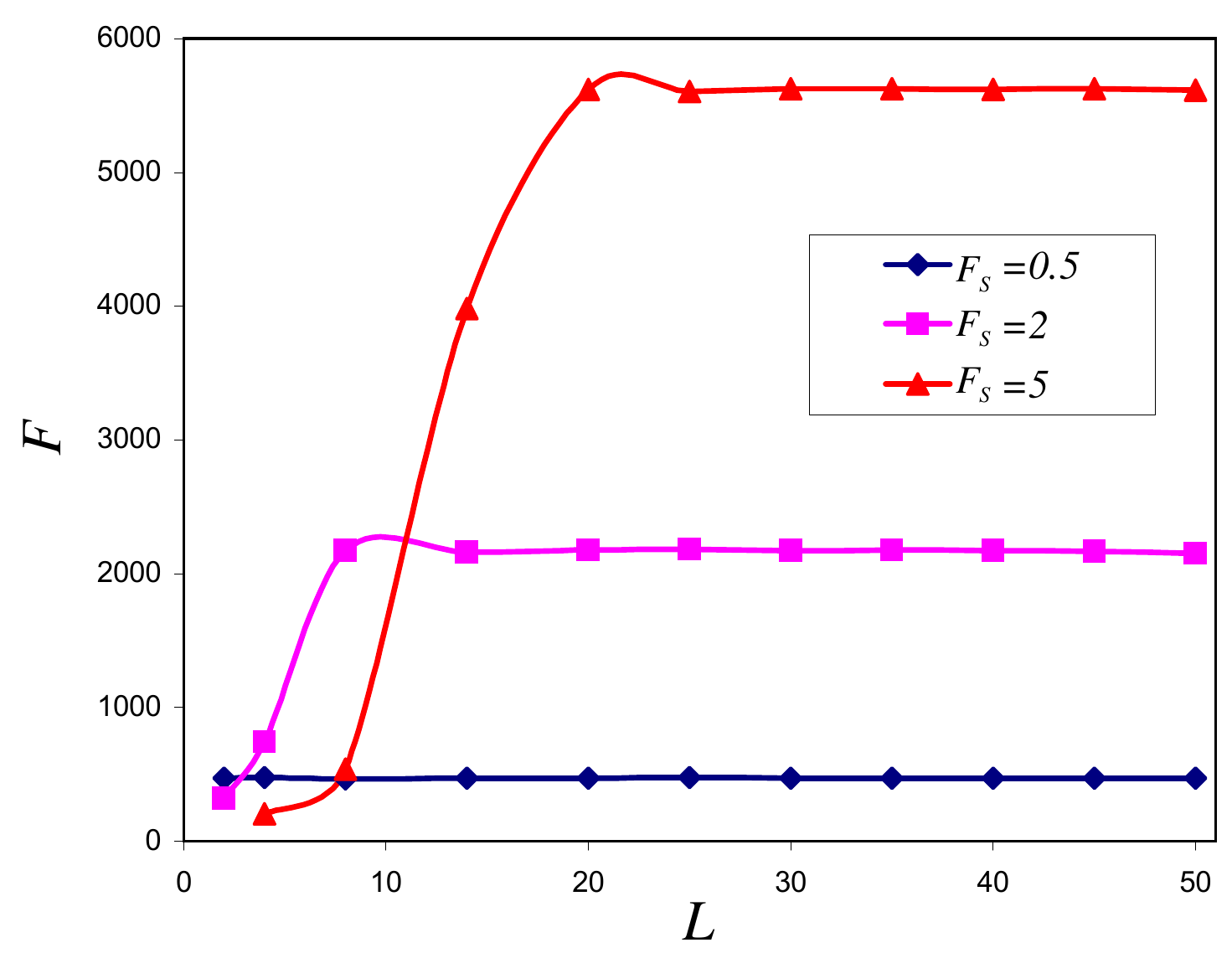}\includegraphics[width=0.48\textwidth]{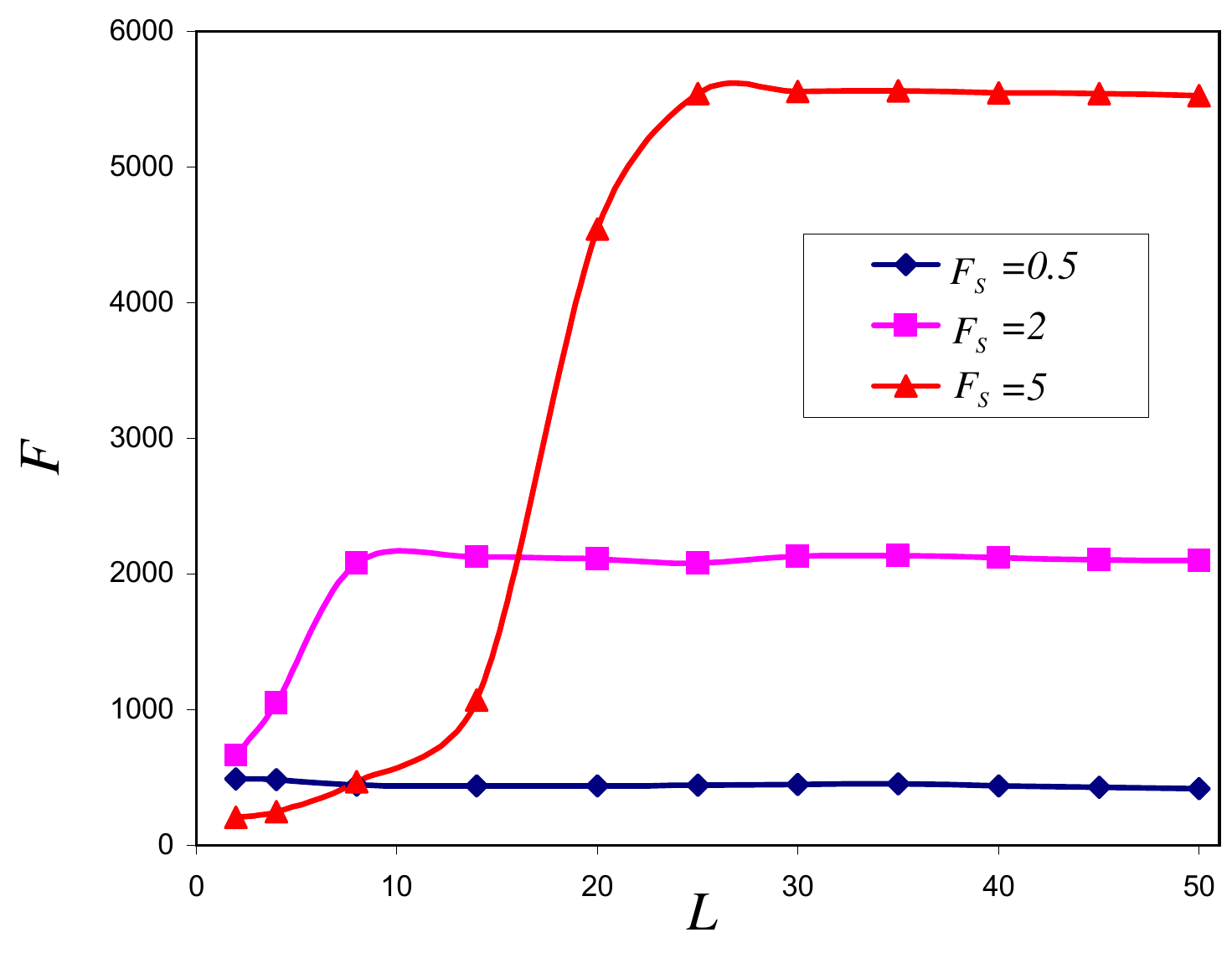}}
\caption{Time-averaged value of the friction force for films with thickness of one (left) and two (right) molecular diameters and different shearing forces as function of load.}
\label{fig9}
\end{figure}

Time averaged friction force as function of load is shown on figure~\ref{fig9}. Here two main regions can be marked out. The first one is linear for low $L$ and it corresponds to the first Amontons' law. The second part is horizontal and is observed for higher loads. Such dependencies can be explained with the use of the ``cobblestone model'' \cite{Dedko2000,GeeML1990,Ruths2005}. Accordingly to this model friction force is defined by two contributions. The first one arises from the presence of the internal adhesion forces between liquid and surface molecules. The second contribution is caused by the externally applied load. Frictional force is defined by the following expression
\begin{equation}
\label{cobblestone}
F=S_{\mathrm{c}}A+CL,
\end{equation}
where constants $S_{\mathrm{c}}$ and $C$ are critical shear stress and the friction coefficient, $A$ is the area of contact and $L$ is the external load. Quantity $S_{\mathrm{c}}$ depends on the adhesion interactions of the film with surfaces. The friction coefficient $C$ is related to the atomic granularity of the surfaces and on the size, shape, and configuration of the liquid molecules in the gap. In general, the smoother the surfaces the smaller should be the value of $C$.

Dependencies on figure~\ref{fig9} may be explained in the following way. As the surfaces in simulations are smooth and water molecules are small and have a simple shape we can assume that frictional force is defined only by the adhesional contribution. Therefore we can consider that $C\approx0$ and in (\ref{cobblestone}) there is only the first term in the right part of the equation. With the increase of load the number and strength of the adhesion bonds of water molecules with surfaces at first rapidly increase. This causes the increase of $S_{\mathrm{c}}$ and from (\ref{cobblestone}) follows the increase of the friction force with $L$ shown on figure~\ref{fig9}. After reaching some external load for the current shearing force some stationary values of the number and strength of adhesion bonds are established. With the increase of $L$ the contact area does not increase, since the model does not take into account the deformation of surfaces. So, the first term in (\ref{cobblestone}) (and therefore friction force) remains constant. The same behaviour exhibits the shear stress $S=F/A$ that is widely used in experiments. Note that in experiments for simple spherical molecules similar dependencies of shear stress $S$ on load $L$ have been obtained~\cite{GeeML1990}. But for real surfaces the constancy of $S$ for high $L$ is caused by the proportional increase of $F$ and $A$ because of the deformation of surfaces. In contrast in simulations the constancy of $S$ is the consequence of the constancy of both $F$ and $A$.

Thus, considered model gives results in many ways corresponding to simple spherical molecules and reproduces for water under low loads the experimentally observed first Amontontons' law. Nevertheless, water molecules cannot be referred to as simple spherical. So, for obtaining more realistic results the model should be improved. For example, the elasticity, roughness, and mutual interactions of surfaces should be taken into account. Such more complicated models are the subject of the further investigations.

\section{Conclusion}

The main result of simulations is the transition of the ultrathin water film to a solidlike state. It is manifested in the decreasing of the diffusion constant and in the ordering of molecules. One layer of molecules between flat rigid diamond plates behaves like simple LJ boundary liquid lubricant and there is no horizontal ordering in it. For two-layer thick film the layering and in-plane ordering are observed. The presence of a driving force causes more rapid forming of ordered structures and the shear ordering is observed. Such behaviour can be explained by the peculiarities of the model, in particular by the presence of the Coulomb interactions between water molecules. Time dependencies of friction force in general reflect a solidlike state of the film and the stick-slip motion is observed for the thicker film under high shear force. But significant difference of time scales in simulations and experiments prevents the definite comparison of corresponding results. The dependencies of time averaged values of friction force and shear stress are similar to experimentally obtained for spherical molecules. They can be explained with the ``cobblestone'' model in the approximation of the adhesion forces domination.

\section{Acknowledgements}

We wish to express our gratitude to the Fundamental Researches State Fund of Ukraine (Grant $\Phi$25.2/013) for supporting of the present work and to L.S.~Metlov for helpful discussions.


\end{document}